\documentclass[epj]{svjour}

\usepackage{graphics}
\usepackage{epsfig}
\usepackage{amsmath}
\usepackage{amssymb}
\newcommand{\bmath}[1]{ \mbox{\boldmath $#1$}  }

\begin{document}

\title{Direct and sequential radiative three-body reaction rates at
  low temperatures}

\author{E. Garrido\inst{1}, R. de Diego\inst{1}, D.V. Fedorov\inst{2} \and A.S.~Jensen\inst{2}} 
\institute{Instituto de Estructura de la Materia, CSIC, Serrano 123, E-28006 Madrid, Spain \and 
           Department of Physics and Astronomy, Aarhus University, DK-8000 Aarhus C, Denmark }

\date{Received: date / Revised version: date}

\abstract{
We investigate the low-temperature reaction rates for radiative
capture processes of three particles.  We compare direct and
sequential capture mechanisms and rates using realistic phenomenological 
parametrizations of the corresponding photodissociation cross sections. 
Energy conservation prohibits
sequential capture for energies smaller than that of the intermediate
two-body structure.  A finite width or a finite temperature allows this
capture mechanism. We study generic effects of positions and widths of
two- and three-body resonances for very low temperatures.  We focus on
nuclear reactions relevant for astrophysics, and we illustrate with
realistic estimates for the $\alpha$-$\alpha$-$\alpha$ and
$\alpha$-$\alpha$-$n$ radiative capture processes.  The direct capture
mechanism leads to reaction rates which for temperatures smaller than
0.1 GK can be several orders of magnitude larger than those of the
NACRE compilation.
}

\authorrunning{E. Garrido et al.}
\maketitle

\section{Introduction}

The techniques employed to solve few-body problems have improved
substantially over the last decade. This is not only due to the computer
development but also to the higher efficiency of both the theoretical formulations 
and the numerical methods.  In particular, a
number of three-body problems are now all solvable at least within a
certain requirement for the accuracy.  The easiest of these problems
are bound states \cite{zhu93,nie01} and perhaps resonances
\cite{cso94,aoy06,alv07a,gar09d,alv10a,die07}, whereas the more challenging are
continuum structures in general \cite{myo01,des06,dan07}.  The most demanding are
reactions where the structures before and after the process are
different \cite{die10,alv08c,fyn09a}.

Applications to nuclear astrophysics have been in demand for
years \cite{fow67,ang99,apr05,muk05,gri06,die10b}. In particular, radiative nuclear three-body capture
processes are indispensable in bridging the $A=5,8$ gaps \cite{apr05,efr96,obe01,fed10}, but
also building up the heavier nuclei may involve three-body reactions.
Also, to pass the waiting points in the rapid proton capture process
necessarily involves three particles, i.e. two protons and a core \cite{gio02,gar04,gri06}. 
Along or outside the neutron dripline, or perhaps
along the rapid neutron capture process, three particles may be
involved as well in decay, capture, or reaction processes \cite{die10,die07,mey94,bar06,die07b}.  

Traditionally, that is mostly for convenience of computation, the
three-body reactions are described as two subsequent two-body
reactions \cite{fow67,efr96,ang99,sum02,gri06,alv08a}.  This two-step sequential mechanism has
proved to be rather efficient in most cases but it cannot always be
accurate. For example, an intermediate configuration of some stability
has to be present, and resonances offer themselves as these stepping
stones. However, they may not be present or they may be too unstable,
such that they decay too quickly after they are formed. It is also
well-known that important contributions arise from the continuum
background to the resonances \cite{myo01,die10b,dan07}.  This is often called the
direct part.  All such difficulties with the sequential models
disappear by using the full three-body formulation of the reactions.
Then the sequential mechanisms are still possible but now as a
contributing process which does not have to be treated separately
\cite{alv07b,alv08a}.

In reactions in stellar environments the temperature, rather than the energy, is often the
parameter. For high temperature, if one of the two-body 
subsystems shows a relatively low-lying narrow resonance, the process
is fairly accurately described by sequential model. In this
case, although the three-body background contribution is missing in the
description of the process, it 
is either small or appropriate parameter adjustments account for this part.
When the intermediate two-body state is missing (or it is too broad), the strength 
function is continuous, without a significant peak structure arising from 
resonances \cite{die10,die07b},  and the
reactions can be described fairly well with knowledge of the
three-body continuum and appropriate energy averaging
\cite{die10,die10b,die07b}.  

At small temperature, where important astrophysical reactions take place, 
only very low energies are involved, which are typically clearly smaller than the energy of the possible 
intermediate two-body states. As a consequence, the situation concerning the reaction mechanism
is unclear. In any case, the complete computation of the direct process in coordinate space is very 
inaccurate due to the fact that a narrow grid of discrete continuum states is required in the very
low energy region. This is often unaffordable from the numerical point of view, and
the problem is usually achieved within the
formalism by assuming a specific form of the strength function or
equivalently of the photodissociation cross section \cite{die10,die10b,die07b}.  
This has the advantage of being a measurable quantity, and thus open to
tests. To circumvent through pure theoretical calculations would
require an accurate model and many points at small energies above the
three-body threshold.

In view of the improved three-body techniques it is surprising that the
sequential model still has not been systematically tested against
genuine three-body computations.  Even though such three-body results
for small energies above threshold are inaccurate, it is still
possible to compare direct and sequential rates by using the different
model formulations with the same common input. This is the purpose of
the present paper which focuses on the low temperature region such that only
three-body energies below the intermediate two-body resonance energies
are relevant for the reaction rate. For these temperatures 
reliability of the sequential model is more than
dubious. 
The idea is to take the established and presently applied formulations for sequential capture, and
compare to a similar formulation which includes the three-body direct capture. These calculations are based
on the assumption of a specific form of the strength function (or the photodissociation cross section), and
therefore do not require the precise numerical solution of the complicated three-body problem.
Even though this is a drastic simplification of the problem, a comparison of the results obtained
with the two possible mechanisms will permit to enlighten the importance of the direct capture
mechanism at low temperature.
In section II we give the appropriate definitions and
formulae for later use. In section III we compute rates in the direct
and sequential models and compare results as a function of the position and
the widths of the two and three-body resonances. In section IV and V we apply
the general method to the triple alpha process and the similar
radiative capture leading to $^9$Be. Finally we summarize and conclude
in section VI.

\section{Direct versus sequential rates}

A method computing the reaction rates for a three-body radiative capture
reaction without assumptions about the underlying mechanism was employed 
previously in \cite{die10,die10b,die07b}. In the first part of this section
we give a simplified description of this method, which, in particular, should 
be used when a direct capture takes place. However, when describing these
processes, the corresponding reaction rates are traditionally described by a 
sequential mechanism via an intermediate structure \cite{fow67,ang99,sum02,efr96}.
The main aspects of this description are summarized in the second part of this
section. By use of the different expressions given, a numerical comparison between 
both descriptions is allowed, in particular on the very low temperature region.
This is done by use of the energy dependent penetration factors given in the
last part of this section for each of the two possible capture mechanisms.

\subsection{The direct picture}

A three-body radiative capture process is usually denoted by
$a+b+c\rightarrow A+ \gamma$, and the reversed reaction is $A+ \gamma
\rightarrow a+b+c$, where $a,b$ and $c$ denote the three particles.
The reaction rate at a given three-body kinetic energy $E$, $R_{abc}(E)$, can 
be obtained by use of Eqs.(20) and (32) in Ref.~\cite{fow67}, which lead to:  
\begin{equation}  
R_{a b c}(E)=\nu!\; \frac{\hbar^3}{c^2} \frac{8\pi}{(\mu_{ab} \mu_{ab,c})^{3/2}} \frac{2g_A}{g_a g_b g_c }
\left( \frac{E_\gamma}{E} \right)^2 
\sigma_\gamma(E_\gamma) \; ,
\label{eq1} 
\end{equation}
where $E_\gamma=E+|B|$ is the photon energy, $B$ ($<0$) is the three-body energy of the
nucleus $A$, $\mu_{a b}$ and $\mu_{a b,c}$ are the reduced masses of
the $a$-$b$ two-body system (related to the Jacobi coordinate
$\bmath{x}$) and the $ab$-$c$ system (related to the Jacobi coordinate
$\bmath{y}$), respectively \cite{nie01}, $g_i=2J_i+1$, where
$J_i$ is the angular momentum of particle $i$ ($i=a,b,c,A$), and $\nu$ is the number of identical
particles in the three-body system.  Finally,
$\sigma_\gamma(E_\gamma)$ is the photodissociation cross section of
the nucleus $A$.

The energy averaged reaction rate is obtained as a function of
the temperature $T$ by using the Maxwell-Boltzmann distribution as
weighting function. For three-particles the Maxwell-Boltzmann distribution 
takes the form:
\begin{equation}
B(E,T)=\frac{1}{2} \frac{E^2}{(k_B T)^3}e^{-\frac{E}{k_BT}},
\end{equation}
where $k_B$ is the Boltzmann constant. We then obtain the following
expression for the energy averaged reaction rate \cite{die10}:
\begin{eqnarray} 
\lefteqn{  \hspace*{-1cm} 
\langle R_{abc}(E) \rangle= \nu!\; \frac{\hbar^3}{c^2} 
\frac{8\pi}{(\mu_{ab} \mu_{ab, c})^{3/2}} \frac{g_A}{g_a g_b g_c}
\times } \nonumber \\ && \times 
\frac{1}{(k_B T)^3} \int_{0}^{\infty} E_\gamma^2 
\sigma_\gamma(E_\gamma)e^{-\frac{E}{k_B T}} dE \;.
\label{eq5} 
\end{eqnarray}
Therefore, once the photodissociation cross section $\sigma_\gamma$ 
for the process $A+\gamma\rightarrow a+b+c$ is known,
the rate $\langle R_{abc}(E) \rangle$ can be easily obtained. For 
instance, it can be obtained directly from the experimental $\sigma_\gamma$
cross section.

Numerical calculation of $\sigma_\gamma$ for the $A+\gamma\rightarrow a+b+c$
process is a delicate task, especially at low temperatures. For instance, the
method used in \cite{die10,die10b,die07b} would require an extremely narrow grid
of discrete continuum states. To achieve this, the continuum wave functions should
be computed in a box of a sufficiently big size, which in turn would imply the need 
of an enormous basis set. All in all, the method becomes inefficient for such low temperatures.
However, when the photodissociation proceeds by populating a three-body
Breit-Wigner shaped resonance of particles $a$, $b$, and $c$ with
total angular momentum $J$ and energy $E_R$, then
$\sigma_\gamma(E_\gamma)$ can be written as:
\begin{equation}
\sigma_\gamma(E_\gamma)=\frac{2J+1}{2g_A}\frac{\pi \hbar^2 c^2}{E_\gamma^2}
\frac{\Gamma_{abc}(E)\Gamma_\gamma(E)}{(E-E_R)^2+\Gamma(E)^2/4},
\label{bweq}
\end{equation}
where $\Gamma_\gamma$ and $\Gamma_{abc}$ are the partial
decay widths of the resonance for gamma and three particle emission
respectively, and $\Gamma=\Gamma_{abc}+\Gamma_\gamma$ is the total
width. 
This is analogous to the method used in the usual sequential picture that
is described in the next section. In this way, by using Eq.(\ref{bweq}) the results
provided by the two different schemes can be easily compared.

It is important to note that Eq.(\ref{eq5}) has been derived without making any assumption 
concerning the reaction mechanism leading to the formation of the nucleus $A$. It is therefore
completely general. In particular, this is the expression to be used when the capture mechanism
is direct, meaning that the capture takes place without populating any intermediate two-body state. 
In this case the width $\Gamma_{abc}(E)$ corresponds to the width for direct decay of the three-body
resonance into particles $a$, $b$, and $c$. For this reason we shall refer to Eq.(\ref{eq5}) as the 
reaction rate in the ``direct" or ``three-body'' picture.

\subsection{The sequential picture}

Let us now assume that one of the internal two-body subsystems, for
instance the $a$-$b$ system, shows a relatively narrow two-body
resonance at some energy $E_r$.  The usual procedure to understand
reactions of the type $a+b+c \rightarrow A+\gamma$ is then to
interpret it as two sequential two-body processes \cite{ang99,sum02}.
In the first step particle $a$ captures $b$ to populate the
intermediate $a$-$b$ two-body resonant state.  In the second step, the
$a$-$b$ system is able, before decaying, to capture particle $c$,
populate some three-body resonance of the nucleus $A$, and then decay
by photo emission into one of the bound states of $A$.

The reaction rate for such a two-step process is given by the rate for
the capture of $c$ by the two-body subsystem $a$-$b$, $\langle
R_{ab,c}(E'',E^\prime) \rangle$, weighted with the rate for formation
of $a$-$b$ \cite{ang99}:
\begin{eqnarray}
\lefteqn{
\langle R_{abc}(E'',E^\prime)\rangle= \frac{\nu!}{1+\delta_{ab}} \frac{8 \pi\hbar}{\mu_{ab}^2}
\left(\frac{\mu_{ab}}{2\pi k_BT} \right)^{3/2}  } \label{eq6} \\ & & 
\int_0^\infty \frac{\sigma_{ab}(E'')}{\Gamma_{ab}(E'')}e^{-E''/k_BT} \langle 
R_{ab,c}(E'',E^\prime) \rangle E'' dE'', \nonumber
\end{eqnarray}
where the total three-body energy, $E = E^\prime + E''$, is given in
terms of the relative energy, $E''$, between particles $a$ and $b$ and
the energy, $E^\prime$, of particle $c$ relative to the center of mass
of $a$-$b$.  The function $\delta_{ab}$ is 1 if $a$ and $b$ are
identical particles, and 0 otherwise.  

The elastic $a$-$b$ cross section, $\sigma_{ab}$, in the equation
above takes the form:
\begin{small}
\begin{equation}
\sigma_{ab}(E'')=(1+\delta_{ab})\frac{g_{ab}}{g_a g_b} 
\frac{\pi}{\kappa^2}\frac{\Gamma_{ab}(E'')^2}{(E''-E_r)^2+(\Gamma_{ab}(E''))^2/4},
\label{eq7}
\end{equation}
\end{small}
where $\kappa^2=2\mu_{ab}E''/\hbar^2$, $E_r$ is the energy of the
resonance in the $a$-$b$ system and $\Gamma_{ab}(E'')$ is the corresponding width, 
and $g_{ab}=2J_{ab}+1$, with $J_{ab}$ being the angular momentum of the
two-body resonance.

Also, following Ref.\cite{ang99}, we have that
\begin{eqnarray}
\lefteqn{\hspace*{-1.5cm} 
\langle R_{ab,c}(E'',E^\prime)\rangle= \frac{8 \pi}{\mu_{ab,c}^2}
\left(\frac{\mu_{ab,c}}{2\pi k_BT} \right)^{3/2}  } \nonumber \\ & & 
\times \int_0^\infty \sigma_{ab,c}(E'',E^\prime)e^{-E^\prime/k_BT} E^\prime dE^\prime, \label{eq11}
\end{eqnarray}
where the cross section $\sigma_{ab,c}(E'',E^\prime)$ for the capture
of particle $c$ by the two-body subsystem $a$-$b$ is related to the
photodissociation cross section through the detailed-balance (or
reciprocity) theorem, which leads to \cite{sum02}:
\begin{equation}
\sigma_{ab,c}(E^\prime)= \frac{g_A}{g_{ab} g_c} \frac{1}{\mu_{ab,c}\; c^ 2} 
       \frac{E_\gamma^2}{E^\prime} \sigma_\gamma(E_\gamma) \;,
\label{bal}
\end{equation}
where $E_\gamma=E+|B|$.

Similarly to the direct process, the photodissociation
cross section $\sigma_\gamma$ again takes the form:
\begin{equation}
\sigma_\gamma(E_\gamma)=\frac{2J+1}{2g_A}\frac{\pi\hbar^2c^2}{E^2_\gamma}
\frac{\Gamma_{ab,c}(E^\prime)\Gamma_\gamma(E^\prime+E'')}{(E-E_R)^2+\Gamma(E',E'')^2/4}.
\label{bwseq}
\end{equation}
This expression is formally identical to Eq.(\ref{bweq}), but now
$\Gamma_{ab,c}$ refers explicitly to the partial width for decay of
the three-body resonance with angular momentum $J$ into the two-body
resonance $a$-$b$ plus particle $c$.  The assumption is that no other
decay mode exists, and the direct decay circumventing this two-body
path is not allowed, or at least negligibly small.  As before,
$\Gamma_\gamma$ is then the partial width for gamma decay, and
$\Gamma=\Gamma_{ab,c}+\Gamma_\gamma$ is the total width.

Replacement of Eq.(\ref{bal}) into (\ref{eq11}), and of
Eq.(\ref{eq11}) into (\ref{eq6}) leads then to the following
expression for the energy averaged reaction rate:
\begin{eqnarray}
\lefteqn{ \hspace*{-0.8cm}
\langle R_{abc}(E'',E^\prime)\rangle= \frac{\nu!}{1+\delta_{ab}} \frac{g_A}{g_{ab}g_c} 
\frac{8 \hbar}{\pi c^2 (k_BT)^3} \frac{1}{\mu_{ab}^{1/2} \mu_{ab,c}^{3/2}} \times } \nonumber \\ &&
\hspace*{-1cm} \times
\int_0^\infty E'' \frac{\sigma_{ab}(E'')}{\Gamma_{ab}(E'')} dE''
\int_{E''}^\infty E_\gamma^2 \sigma_\gamma(E_\gamma)e^{-E/k_BT} dE 
\label{fullseq}
\end{eqnarray}
where we have replaced the dependence on $E^\prime$ by $E$ by use of
$E=E^\prime+E''$.  We shall refer to the expression in
Eq.(\ref{fullseq}) as the reaction rate in the ``sequential'' picture.

\subsubsection{Extreme sequential limits}
In this subsection we shall derive the form that the reaction rate (\ref{fullseq}) takes in the extreme
cases of an infinitely narrow three-body resonance and/or an infinitely narrow intermediate two-body state. We shall
refer to these rates as the ones obtained in the ``extreme sequential'' picture.

The connection between the reaction rates (\ref{eq5}) and (\ref{fullseq}) is made evident for the particular
case of a very narrow $a$-$b$ two-body resonance ($\Gamma_{ab}$ very small compared to
the energy of the resonance). In this case we have that  Eq.(\ref{eq7}) can be replaced by:
\begin{equation}
\frac{\sigma_{ab}(E'')}{\Gamma_{ab}(E'')}=
(1+\delta_{ab}) \frac{g_{ab}}{g_a g_b}
\frac{2\pi^2}{\kappa^2} \delta(E''-E_r),
\label{eq9}
\end{equation}
from which Eq.(\ref{fullseq}) becomes:
\begin{eqnarray} 
\lefteqn{  \hspace*{-1cm} 
\langle R_{abc}(E) \rangle= \nu!\; \frac{\hbar^3}{c^2} 
\frac{8\pi}{(\mu_{ab} \mu_{ab, c})^{3/2}} \frac{g_A}{g_a g_b g_c}
\times } \nonumber \\ && \times 
\frac{1}{(k_B T)^3} \int_{E_r}^{\infty} E_\gamma^2 
\sigma_\gamma(E_\gamma)e^{-\frac{E}{k_B T}} dE \;.
\label{seq} 
\end{eqnarray}
and the three-body energy $E$ is given by $E_r+E^\prime$. 

We can immediately see that the reaction rates in Eqs.(\ref{eq5}) and
(\ref{seq}) are identical except for the lower limit in the
integration. From $0$ in the first case and from the intermediate
two-body resonance energy $E_r$ in the second. This reflects the fact
that in the extreme sequential picture described by Eq.(\ref{seq}),
where an infinitely narrow intermediate two-body resonance is assumed
to be populated, the total three-body energy $E$ must be larger than
$E_r$.  For three-body energies smaller than $E_r$ this sequential
picture cannot provide any decay, since the energy is too small to
populate the two-body intermediate resonance. This means that in this 
extreme limit, when the three-body resonance energy $E_R$ is smaller than $E_r$, the
sequential decay has to proceed through the tail of the cross section in Eq.(\ref{bwseq})
corresponding to $E>E_r>E_R$.

We can also consider the case where the width $\Gamma_{ab,c}$ in
Eq.(\ref{bwseq}) is very small. Assuming that $\Gamma_\gamma \lll
\Gamma_{ab,c}$ is still fulfilled, we then have from Eq.(\ref{bwseq})
that:
\begin{equation}
\sigma_\gamma(E_\gamma)=\frac{2J+1}{g_A} \frac{\Gamma_\gamma(E) \pi^2 \hbar^2 c^2}{E_\gamma^2}
                        \delta(E-E_R),
\label{3bnarrow}
\end{equation}
which leads to
\begin{eqnarray}
&&
\langle R_{abc}(E) \rangle = \nu! \frac{2J+1}{g_a g_b g_c} \frac{(2\pi)^2\hbar^5}{(k_B T)^3} 
\frac{\Gamma_\gamma(E_R)}{(\mu_{ab}\mu_{ab,c})^{3/2}} e^{-E_R/k_B T} \times 
    \nonumber \\ &&
\times \int_0^{E_R} \frac{\Gamma_{ab}(E'')}{(E''-E_r)^2+ (\Gamma_{ab}(E''))^2/4} dE''. 
\label{lim2}
\end{eqnarray} 
From the expression above we can see that when $E_R<E_r$ the decay
in this limit must proceed through the tail of the cross section in
Eq.(\ref{eq7}) corresponding to $E''<E_R<E_r$.

When the widths $\Gamma_{ab,c}$ and $\Gamma_{ab}$ are both very small then the approximations 
(\ref{eq9}) and (\ref{3bnarrow}) can be used  simultaneously, and either 
Eq.(\ref{seq}) or (\ref{lim2}) lead to the following expression for the
energy averaged reaction rate:
\begin{equation}
\langle R_{abc}(E)\rangle = \nu! \frac{2J+1}{g_a g_b g_c} \frac{(2\pi)^3\hbar^5 }{(k_B T)^3} 
\frac{\Gamma_\gamma(E_R)}{(\mu_{ab}\mu_{ab,c})^{3/2}} e^{-E_R/k_B T}
\label{eqlim}
\end{equation}
which is valid when the two-body resonance energy $E_r$ is below the
three-body resonance energy $E_R$. Otherwise, if $E_R<E_r$ the
reaction rate is zero in this limit of very small widths.  The
expression in Eq.(\ref{eqlim}) does not depend on the energy $E_r$ of
the intermediate two-body resonance, and it is actually the same
expression that one gets when the approximation (\ref{3bnarrow}) is
used on the reaction rate (\ref{eq5}) obtained in the direct picture.

At this point it is important to remind ourselves that Eqs.(\ref{seq}),
(\ref{lim2}), and (\ref{eqlim}) are limiting cases involving very, or
even infinitely, narrow resonances. In particular, Eq.(\ref{seq})
permits to see very easily the connection between Eqs.(\ref{eq5}) and
(\ref{fullseq}). However, correct sequential calculations have to be
performed by use of Eq.(\ref{fullseq}), as done for instance in
\cite{ang99} or \cite{sum02}.

\subsection{Low energy dependence}

The Boltzmann exponents in Eqs.(\ref{eq5}) and (\ref{fullseq}) only
allow contribution from three-body energies up to a few times $k_BT$.
For instance, for a typical temperature in the core of a star of
$10^7$ K we have that $k_B T\approx 10^{-3}$ MeV, and only energies of
the order of a few keV's and below can contribute to the rate.  These
energies are very small at the nuclear scale, where excitation
energies are typically of the order of the MeV. Therefore, for such
low temperatures, the reaction rate in Eq.(\ref{eq5}) or
(\ref{fullseq}) is dictated by the low energy tail of the
corresponding cross sections.

As seen in Eqs.(\ref{bweq}), (\ref{eq7}), and (\ref{bwseq}) the
behavior of the cross sections at very low energies is determined by
the energy dependence of the partial widths, which are proportional to the 
penetration factor through the
barrier responsible for the corresponding resonance ($\Gamma(E) \propto
P(E)$).  The constant of proportionality is determined by assuming
that when evaluated at the resonance energy $E_{res}$ the width
$\Gamma(E_{res})$ provides the established (or experimental) width of
the resonance $\Gamma_0$. This is therefore leading to the expression:
\begin{equation}
\Gamma(E)= \Gamma_0 \frac{P(E)}{P(E_{res})},
\label{pfactor}
\end{equation}
where the explicit form of the penetration factor $P(E)$ should be determined for
each particular case.

For gamma decay it is well known from text books (see for instance the
appendix E.1 of \cite{pre75}) that the decay constant, and therefore
the penetration factor and $\Gamma_\gamma(E)$, is proportional to
$E_\gamma^{2\lambda+1}$, where $E_\gamma=E+|B|$ is the photon energy
and $\lambda$ is the multipolarity of the electromagnetic
transition. As discussed in Eq.(\ref{pfactor}), the proportionality
constant is determined in such a way that $\Gamma_\gamma(E_R)$ is
equal to the experimental value $\Gamma_\gamma$ for the gamma decay
width. The parametrization is then:
\begin{equation}
\Gamma_\gamma(E)=\Gamma_\gamma \left( \frac{E+|B|}{E_R+|B|}\right)^{2\lambda+1}.
\label{gamwidth}
\end{equation} 

In the same way, when the Coulomb interaction is not involved, the
partial widths $\Gamma_{ab,c}(E')$ and $\Gamma_{ab}(E'')$ in
Eqs.(\ref{eq7}) and (\ref{bwseq}) for the sequential picture take the
form:
\begin{eqnarray}
\Gamma_{ab,c}(E')&=&\Gamma_{ab,c}\left( \frac{E'}{E_R-E_r}\right)^{\ell_{ab,c}+1/2}
\label{gamseq1}  \\
\Gamma_{ab}(E'')&=&\Gamma_{ab}\left( \frac{E''}{E_r}\right)^{\ell_{ab}+1/2},
\label{gamseq2}
\end{eqnarray}
where $\ell_{ab,c}$ and $\ell_{ab}$ are the relative orbital angular momenta between particle $c$
and the center of mass of $a$-$b$ and between particles $a$ and $b$, respectively, and where 
we have taken into account that the energy dependence of these partial widths is given by Eqs.(22) and
(23) of Ref.\cite{gar05}.

When the Coulomb interaction is decisive, insertion of Eqs.(10) and
(11) into (3) of Ref.\cite{gar05} leads to the following energy
dependence of the partial widths in the sequential picture:
\begin{eqnarray}
\Gamma_{ab,c}(E')&=&\Gamma_{ab,c}\frac{1+e^{2b_{ab,c}/\sqrt{E_R-E_r}}}{1+e^{2b_{ab,c}/\sqrt{E'}}}
\label{gamseq3}  \\
\Gamma_{ab}(E'')&=&\Gamma_{ab}\frac{1+e^{2b_{ab}/\sqrt{E_r}}}{1+e^{2b_{ab}/\sqrt{E''}}},
\label{gamseq4}
\end{eqnarray}
where
\begin{eqnarray}
b_{ab,c}&=&\frac{\pi}{2}(Z_a+Z_b)Z_c e^2 \sqrt{\frac{2\mu_{ab,c}}{\hbar^2}}
\label{babc}
 \\
b_{ab}&=&\frac{\pi}{2} Z_a Z_b e^2 \sqrt{\frac{2\mu_{ab}}{\hbar^2}},
\label{bab} 
\end{eqnarray}
and where $Z_a$, $Z_b$, and $Z_c$ are the charges of particles $a$, $b$, and $c$, respectively,
and $e$ is the electron charge. 

For direct decay we get, from Eqs.(12) and (3) of \cite{gar05},
for Coulomb potentials that:
\begin{equation}
\Gamma_{abc}(E)=\Gamma_{abc}\frac{1+e^{2b_{abc}/\sqrt{E_R}}}{1+e^{2b_{abc}/\sqrt{E}}}
\label{dir2}
\end{equation}
with
\begin{small}
\begin{equation}
b_{abc}=\frac{\pi}{2} \sqrt{\frac{2}{\hbar^2(m_a+m_b+m_c)}}
\left( \sum
\left( Z_i Z_j e^2\right)^{2/3} \left( m_i m_j\right)^{1/3}
\right)^{3/2}
\label{bdir}
\end{equation}
\end{small}
where $m_a$, $m_b$, and $m_c$ are the masses of the particles and the
sum runs over the three possible pairs of particles.

For direct decay, without Coulomb potentials, we get the energy
dependence of $\Gamma_{abc}(E)$ in Eqs.(\ref{bweq}) and (\ref{eq5})
from Eq.(21) of Ref.\cite{gar05}:
\begin{equation}
\Gamma_{abc}(E)=\Gamma_{abc}\left( \frac{E}{E_R} \right)^{\ell_{ab,c}+\ell_{ab}+2}.
\label{dir1}
\end{equation}

\section{Parameter dependence}

The computed reaction rates depend on the choice for the reaction  mechanism,
i.e. either direct or sequential. To illustrate it we start with a system
of three identical charged particles.  We shall assume that this
three-body system has a resonance at the energy  $E_R=0.15$ MeV
and $\Gamma_\gamma=10^{-9}$ MeV. All the degeneracy factors $g_i$ are
taken equal to 1, the charges are taken twice the proton charge, and
the masses equal to the mass of the alpha particle.

If the capture process is assumed to proceed directly, the reaction
rate is then given by Eqs.(\ref{bweq}) and (\ref{eq5}), and for three
charged particles, the energy dependence of the width for particle
decay is given by Eqs.(\ref{dir2}) and (\ref{bdir}). On the other
hand, if we assume that the capture process takes place sequentially
through an intermediate two-body resonance, the reaction rate is
instead given by Eqs. (\ref{eq7}), (\ref{bwseq}), and (\ref{fullseq}),
and the energy dependence of the widths, $\Gamma_{ab}$ and
$\Gamma_{ab,c}$, for particle decay in Eqs.(\ref{eq7}) and
(\ref{bwseq}) are now as in (\ref{gamseq4}) and (\ref{gamseq3}).

\begin{figure}[t!]
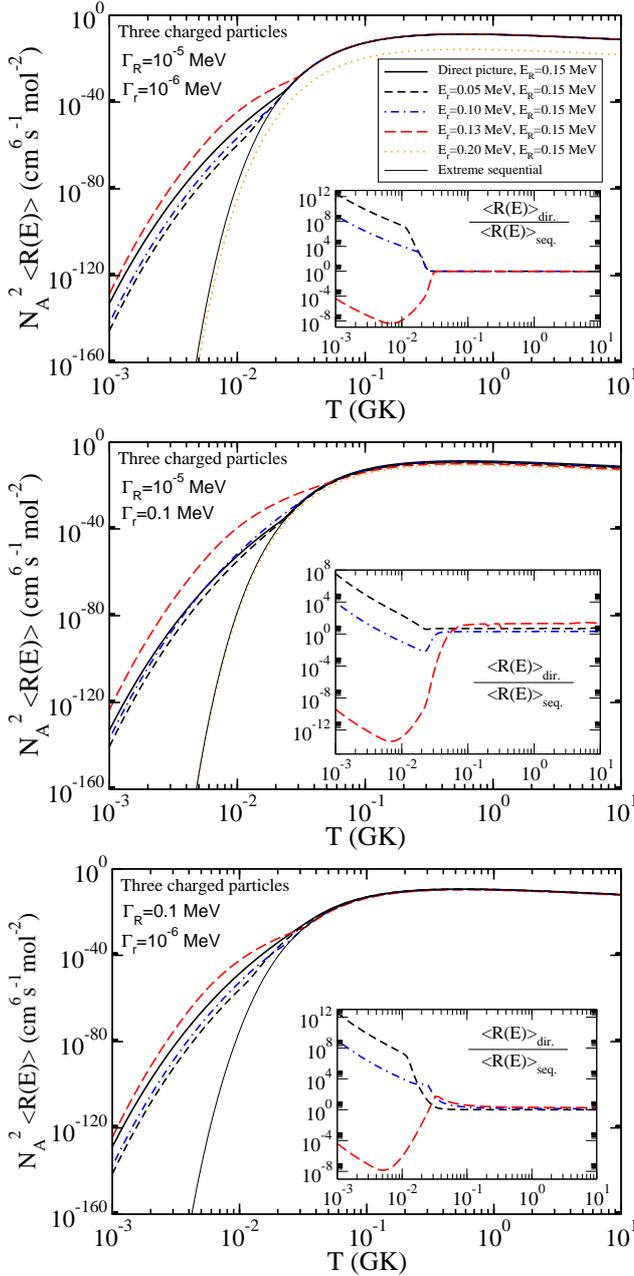

\psfig{figure=schem1.eps,width=8.3cm,angle=0}
\psfig{figure=schem1-broad2.eps,width=8.3cm,angle=0}
\psfig{figure=schem1-broad.eps,width=8.3cm,angle=0}
\vspace*{-1mm}
\caption{(color online) Reaction rates for the radiative capture of three identical
charged particles (with charge and mass equal to the ones of the $\alpha$-particle). 
The widths $\Gamma_R$ and $\Gamma_r$ of the three- and two-body resonances are
10$^{-5}$ MeV and 10$^{-6}$ MeV, respectively in the upper part, 10$^{-5}$ MeV and 
0.1 MeV, respectively in the central part, and 0.1 MeV and 10$^{-6}$ MeV, respectively 
in the lower part.  The thick-solid curve shows the reaction rate in the direct picture. 
The short-dashed, dot-dashed,
and long-dashed curves are the rates in the sequential picture for two-body energies $E_r=0.05$ MeV, 
0.10 MeV, and 0.13 MeV. The ratio between the rates in the direct and sequential pictures
are shown in the insets. The thin solid line is the rate in the extreme 
sequential limit. The dotted curve is the reaction rate obtained when $E_r=0.20$ MeV $(E_R<E_r)$ (see text).
$N_A$ is the Avogadro's number.}
\label{scheme}
\end{figure}

\begin{figure}[t!]
\hspace{-1cm}
\psfig{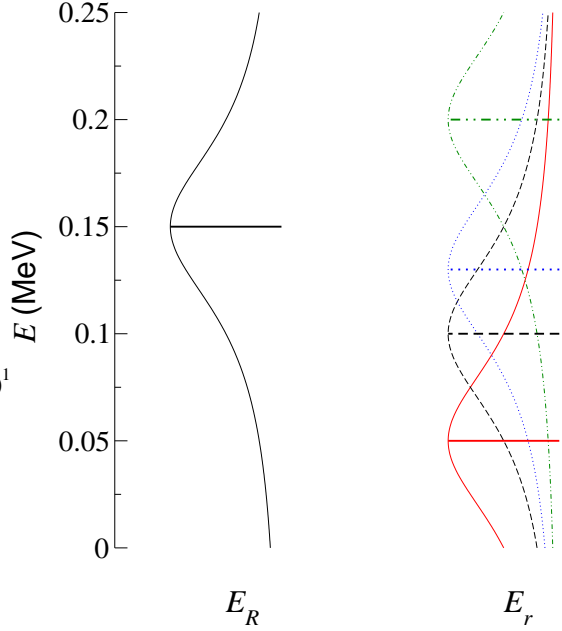}
\caption{(color online) Energy diagram showing the three-body resonance energy $E_R$ and 
two-body resonance energies $E_r$ used in the schematic calculations in section III. The curves
are the lorentzians corresponding the resonances when having a width of 0.1 MeV.}
\label{diag}
\end{figure}

As a first case, let us assume that the three-body and the
intermediate two-body resonances are both narrow, e.g. by choosing
$\Gamma_R=10^{-5}$ MeV and $\Gamma_r=10^{-6}$ MeV. The computed
reaction rate in the direct picture is then given by the thick solid
line in the upper part of Fig.~\ref{scheme}. For the sequential case,
when the capture proceeds through the intermediate two-body resonance,
as seen in Eqs.(\ref{eq7}) and (\ref{fullseq}), the reaction rate
depends as well on the energy of the two-body resonance. We have chosen
the energies of 0.05 MeV, 0.10 MeV, 0.13 MeV, and 0.20 MeV, as indicated by 
the straight lines in the right part of Fig~\ref{diag}. The straight
line in the left part indicates the energy chosen for the three-body resonance.
In this case of very narrow three-body and two-body resonances, the lorentzians 
describing them can not be distinguished from the lines indicating 
the energies in Fig.~\ref{diag}. The corresponding reaction rates in the
sequential picture are given by the short-dashed, dot-dashed, and long-dashed curves 
in the upper part of Fig.~\ref{scheme} for the three cases where $E_r$ takes 
the values 0.05 MeV, 0.10 MeV, and 0.13 MeV, respectively.

As we can see in the upper part of Fig.~\ref{scheme}, for temperatures below about 0.02
GK the reaction rates in the two pictures differ from each other.  In
the first two cases, $E_r=0.05$ MeV and $E_r=0.10$ MeV, the sequential
rate is below the direct one, while for $E_r=0.13$ MeV, the sequential
rate at low temperatures is above the direct one. This behavior is 
determined by the interplay between $\Gamma_{ab,c}$ and $\Gamma_{ab}$ in 
Eqs.(\ref{gamseq3}) and (\ref{gamseq4}). When $E_r$ increases and approaches
$E_R$, then $\Gamma_{ab,c}$ increases, implying a more significant tail in 
the cross section (\ref{bwseq}), and therefore favoring an increase of the reaction
rate at low temperatures. However, at the same time the increase of $E_r$
decreases $\Gamma_{ab}$, which for the same reason tends to reduce the reaction rate. The final
result of the competition between these two effects pushing in opposite 
directions depends on the masses and charges of the particles involved in the process.
In our case the total effect amounts to an increase of the reaction rate when $E_r$ increases.
The comparison between the direct and the sequential calculations is
better seen in the inner part of the figure, where the different
curves show the ratio between the rates in the direct and sequential
pictures. As we can see, the difference between them reaches up to 12
orders of magnitude for $T=0.001$ GK.

\begin{figure}[t!]
\psfig{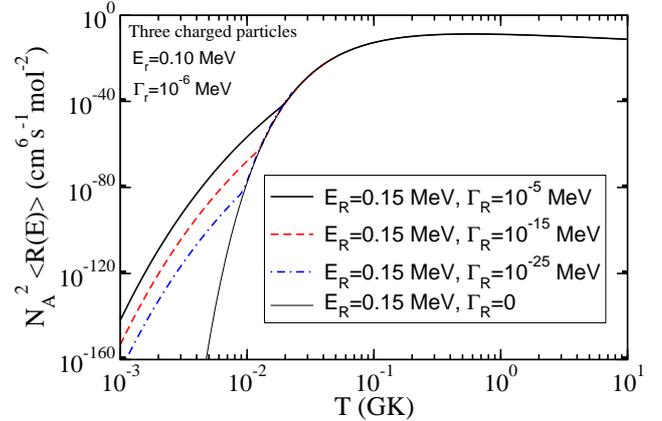}
\caption{(color online) 
Reaction rates in the sequential picture for the same system as in the upper part of Fig.~\ref{scheme}
with $E_r=0.10$ MeV but for different small values for the width of the three-body resonance $\Gamma_R$. 
The thick-solid and thin-solid curves are the same as the dot-dashed and thin-solid curves in the upper 
part of Fig.~\ref{scheme}.
The dashed and dot-dashed curves are calculations with $\Gamma_R=10^{-15}$ MeV and $\Gamma_R=10^{-25}$ MeV,
respectively.  }
\label{wths}
\end{figure}

For temperatures higher than about 0.03 GK the rates obtained in the
sequential picture are not sensitive to the properties of the
intermediate two-body resonance, and the computed rate agrees with the
one obtained in the direct picture. This is consistent with the fact
that when the three-body resonance and the two-body one are both very
narrow the reaction rate can be approximated by Eq.(\ref{eqlim}),
which depends neither on $E_r$ nor on $\Gamma_r$. This is
referred to as the rate in the extreme sequential picture, and
it is shown in the upper part of
Fig.~\ref{scheme} by the thin solid line.  As we can see, for
$T\gtrsim 0.03$ GK this approximation agrees with both the full direct
and the sequential calculations. At low temperatures this is not true
anymore, which shows that the finite width of the resonances influences
the rates, and the limit of zero width leading to Eq.(\ref{eqlim})
becomes increasingly invalid.  The reason is that the effect from the
tails of the resonances, where the penetration factors play a role, is
completely absent in Eq.(\ref{eqlim}).  This is easily recognized
in Fig.~\ref{wths}, where the thick-solid and thin-solid curves are the same as 
the dot-dashed and thin-solid ones in the upper part of Fig.~\ref{scheme} 
(sequential reaction rate with $E_r=0.10$ MeV and reaction rate in the extreme
sequential limit, respectively), while the dashed and dot-dashed curves are the
calculations with $\Gamma_R=10^{-15}$ MeV and $\Gamma_R=10^{-25}$ MeV,
respectively. As we can see, the narrower the three-body resonance the
lower the temperature at which the full sequential calculation matches
the curve in the extreme sequential picture.

The central and lower parts of Fig.~\ref{scheme} show the same reaction
rates as in the upper part, but for $\Gamma_r=0.1$ MeV and
$\Gamma_R=0.1$ MeV, respectively. A feeling of the overlap between the
three-body and two-body resonances in the different cases can be 
got from Fig.~\ref{diag}, where the curves plotted are lorentzians 
with $\Gamma=0.1$ MeV. The behavior observed in the central and lower parts
of Fig.~\ref{scheme} is similar to the one found in the upper part, where both 
resonances are simultaneously narrow. Also the ratio between the direct and
sequential calculations (inner part of the figures) is similar. In
these two cases, the extreme sequential curves correspond to the ones
obtained with the expressions (\ref{lim2}) and (\ref{seq}),
respectively. As we can see, Eq.(\ref{lim2}) does actually depend on
the properties of the two-body resonance energy, which leads in this
case (central part of Fig.~\ref{scheme}) to some small differences
between the different calculations. In particular, the curve in the 
figure (thin solid line) has been obtained using $E_r=0.10$ MeV.

It is important to note that the equations in Ref.\cite{gar05} which
have led to Eqs.(\ref{gamseq1}) and (\ref{gamseq4}) have been obtained
using the adiabatic approximation to describe the three-body system.
The penetration factors are then extracted from the WKB tunneling
probability through the potential barrier of the lowest adiabatic
potential. When one of the two-body systems has a resonance, this
lowest adiabatic potential shows a flat region at the two-body
resonance energy, and eventually goes to zero. The narrower the
two-body resonance, the longer its lifetime, and the larger the flat
region in the adiabatic potential.  In fact, in the limit of zero
width (or infinite lifetime) we actually have a bound two-body state,
and the adiabatic potential goes asymptotically to the two-body energy
of the bound state. This implies that the expressions (\ref{gamseq1})
and (\ref{gamseq3}) are meaningful only when the three-body resonance
energy $E_R$ is bigger than the two-body resonance energy $E_r$. 

In the opposite case, when $E_R<E_r$, the reaction mechanism is
dominantly direct, since the intermediate configuration is
energetically forbidden. However, it is still possible to exploit the
sequential mechanism but obviously not until the decay is completed.
The distance between particles must be limited, or equivalently the
time allowed for the non-energy conserving process is limited. If the
decay still has to take place through the sequential decay mechanism
then the two-body resonance width has to be finite (see the dot-dashed
curve in Fig.~\ref{diag}, where even if it corresponds to
$E_R<E_r$ a significant part of the resonance tail falls in the
energetically allowed region). This mechanism was
previously named virtual sequential decay \cite{gar04}.

From the discussion above we then have that, when $E_R<E_r$, the barrier 
determining the penetration factor is much
bigger than in the opposite case, and in fact the narrower the two-body resonance
the thicker the barrier.  As a consequence, for $E_R<E_r$ we have that
Eqs.(\ref{gamseq1}) and (\ref{gamseq3}) can not be used, and the
three-body width is expected to be much smaller than when $E_R>E_r$.
For this reason, the reaction rate for the cases where $E_R<E_r$ can
only be estimated from Eq.(\ref{lim2}), where the value of $\Gamma_{ab,c}$ has
been assumed to be zero. From this equation, since $E_R<E_r$, we can
immediately see that the reaction rate is produced by the tail of the
intermediate two-body resonance (dot-dashed curve in Fig.~\ref{diag}). 
Clearly, the broader this resonance
the larger the reaction rate. This is shown by the dotted curves in
the upper and central parts of Fig.~\ref{scheme} for $\Gamma_r=10^{-6}$
MeV and $\Gamma_r=0.1$ MeV, respectively. These two calculations have
been made with resonance energies $E_r=0.20$ MeV and $E_R=0.15$ MeV.

In the central part the width of the two-body resonance is not small,
$\Gamma_r = 0.1$~MeV, and the difference with the previous
calculations is hardly seen in logarithmic scale, although it is two
orders of magnitude smaller than the direct picture calculation. When
$\Gamma_r=10^{-6}$ (upper part), the dotted curve is up to eight
orders of magnitude smaller than the other calculations, consistent
with the fact that in the limit where $\Gamma_r=0$ the rate provided
by Eq.(\ref{seq}) is zero. In the lower part of the figure the
corresponding calculation with $E_r=0.20$ MeV is not shown. First,
because for $E_R<E_r$ the width of the three-body resonance is
expected to be very small, which is not the case in this calculation,
and second, because even if we insist in computing it, the penetration
factors (\ref{gamseq1}) and (\ref{gamseq3}) are obviously meaningless
for $E_R<E_r$, and another parametrization would have to be designed.

\begin{figure}[t]
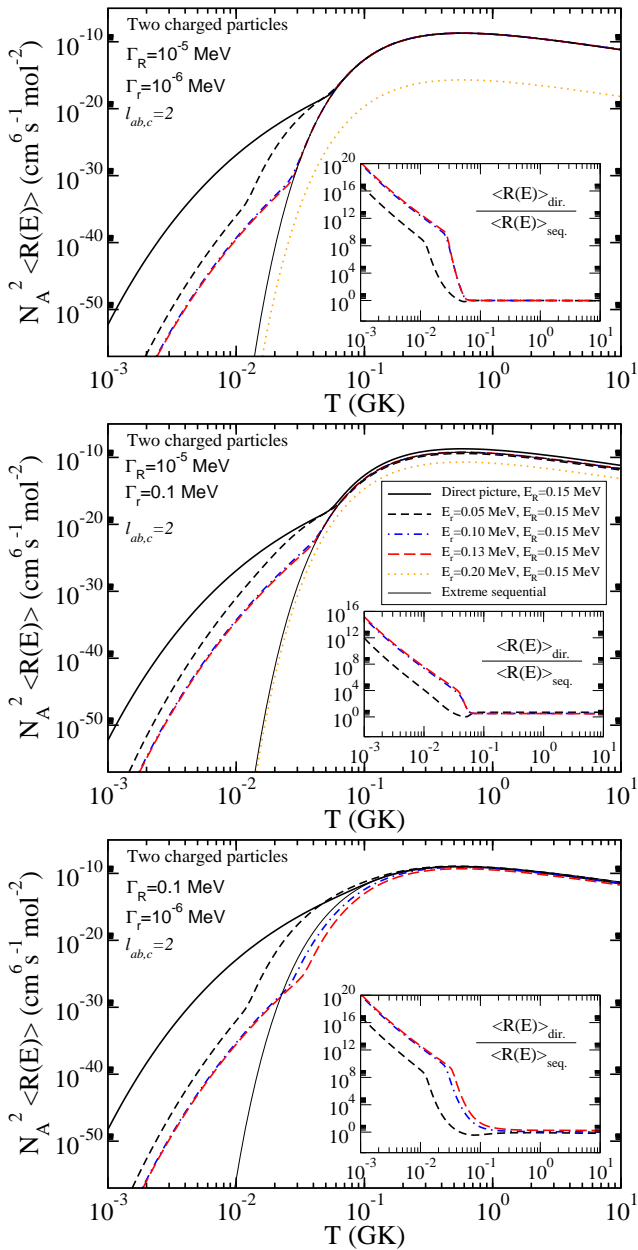

\psfig{figure=schem2.eps,width=8.3cm,angle=0}
\psfig{figure=schem2-broad2.eps,width=8.3cm,angle=0}
\psfig{figure=schem2-broad.eps,width=8.3cm,angle=0}
\caption{(color online) The same as in Fig.~\ref{scheme} when only the
  two particles forming the intermediate two-body resonance are
  charged. The centrifugal barriers correspond to orbital
  angular momentum $\ell_{ab,c}=2$. }
\label{schemb}
\end{figure}

\begin{figure}[t]
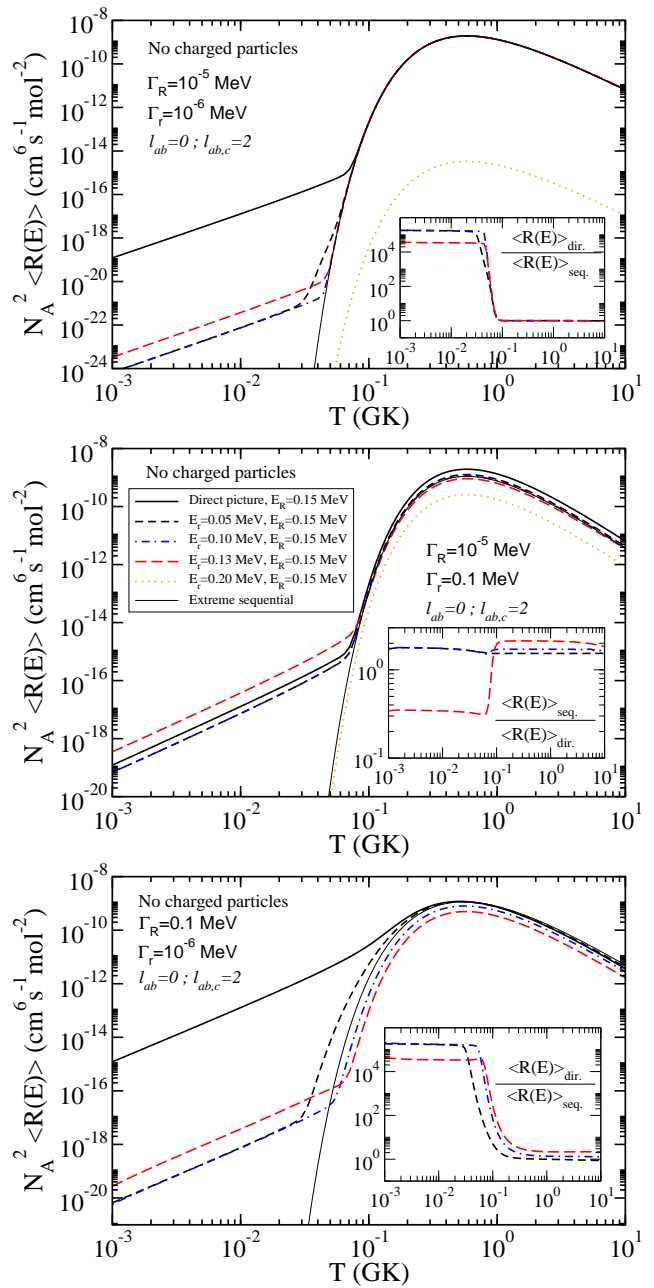

\psfig{figure=schem3.eps,width=8.3cm,angle=0}
\psfig{figure=schem3-broad2.eps,width=8.3cm,angle=0}
\psfig{figure=schem3-broad.eps,width=8.3cm,angle=0}
\caption{(color online) The same as in Fig.~\ref{scheme} for three
  identical uncharged particles.  The centrifugal barriers correspond
  to orbital angular momentum $\ell_{ab,c}=2$. }
\label{schemc}
\end{figure}

For completeness, we show in Figs.~\ref{schemb} and \ref{schemc} the
same calculations as in Fig.~\ref{scheme}, but only for the cases where,
respectively, two or none of the particles forming the two-body
resonance are charged.  The meaning of the curves is as in
Fig.~\ref{scheme}, and the only difference lies in the use of the
proper penetration factors for the cases where the Coulomb interaction
can be neglected. 

In particular, in Fig.~\ref{schemb} we assume that
the two charged particles are the ones forming the intermediate
two-body resonance in the sequential picture. Therefore, the
penetration factors to be used are (\ref{dir2}) in the direct case,
and (\ref{gamseq1}) and (\ref{gamseq4}) in the sequential case. In this
case, when $E_r$ increases and approaches $E_R$ we have that $\Gamma_{ab,c}$
increases as a polynomial, while $\Gamma_{ab}$ decreases exponentially. 
As a consequence, the decrease in the reaction rate due to the decrease of 
$\Gamma_{ab}$ dominates over the increase produced by the increase of 
$\Gamma_{ab,c}$, and the final result is that the computed reaction rate
decreases at low temperatures when $E_r$ increases, although the difference
between the calculations when $E_r=0.10$ MeV and $E_r=0.13$ MeV is hardly seen
in the figure. In Fig.~\ref{schemc} the penetration factors to be used are (\ref{dir1})
in the direct case, and (\ref{gamseq1}) and (\ref{gamseq2}) in the
sequential case.  The general trends of the reaction rates shown in
Figs.~\ref{schemb} and \ref{schemc} are similar to the ones shown and
discussed in Fig.~\ref{scheme}. The main difference is the absolute
value of the rates at small temperatures, which increases enormously
when the Coulomb barrier disappears.

These general features of the reaction rates, especially the ones concerning
their dependence on model assumptions of the capture process, can now
be used in practical cases of (astro)physical interest. We shall in
the next two sections focus on the triple alpha reaction (three
identical charged particles with narrow three-body and intermediate
two-body resonances), 
and on the $\alpha+\alpha+n\rightarrow ^9\mbox{Be}+\gamma$
process (two identical charged particles forming a narrow two-body
resonance, and a relatively broad three-body resonance).

\section{The triple alpha rate}

The reaction rate for the triple alpha reaction, $\alpha+\alpha+\alpha
\rightarrow \mbox{$^{12}$C}+\gamma$, at very low temperatures is quite
controversial. In fact, in the recent work \cite{oga09}, the reaction
rate at a temperature of 10$^{-2}$ GK has been found to be about 20
orders of magnitude bigger than the one given in the NACRE compilation
\cite{ang99}. Such enormous increase in the reaction rate would have
dramatic consequences for the late stages of the stellar evolution in
low mass stars \cite{dot09}.

As demonstrated in the previous sections, the key quantity to obtain
an accurate reaction rate for a radiative capture process is the
photodissociation cross section for the inverse reaction. This
means that in our case $\sigma_\gamma$ for the $\mbox{$^{12}$C}+\gamma
\rightarrow \alpha+\alpha+\alpha$ reaction is needed.  This cross
section is, to a large extent, determined by the resonance spectrum in
$^{12}$C, which is rather well known except for the lowest 2$^+$
resonance, whose excitation energy is not well established yet.
Fortunately, as shown in \cite{die10b}, this resonance is playing a
role in the reaction rate only for temperatures higher than about 2
GK. Therefore, this uncertainty in the $^{12}$C spectrum is irrelevant
for the analysis of the reaction rate at very low temperatures, which
is dominated by the E2 transition $0^+ \rightarrow 2^+_1$
\cite{die10b}. This notation defines the transitions from the $0^+$
continuum states in the three $\alpha$ system to the bound $2^+$ state
in $^{12}$C.

Furthermore, as also shown in \cite{die10b}, for temperatures higher
than 0.1 GK the computed contribution to the reaction rate from the
$0^+ \rightarrow 2^+_1$ transitions is very similar when performing a
full three-body calculation and when assuming a sequential capture
through the low-lying $0^+$ state in $^8$Be (0.092 MeV above the
two-body threshold). This is due to the existence of the narrow $0^+$
Hoyle three-body resonance in $^{12}$C at about 0.38 MeV above
threshold. This state heavily dominates the calculation, which has a
large strength corresponding to strong population and subsequent decay
of the Hoyle state. Three-body calculations have shown that the
particle decay of the Hoyle state proceeds almost fully sequentially
\cite{alv08}, and therefore so does also the inverse process.

However, for very small temperatures the relevant three-body energies
in the reaction rate are clearly below the 0.092 MeV of the
intermediate 0$^+$ resonance in $^8$Be. It is then not so obvious that
the sequential picture through that $0^+$ state is still
appropriate. The occurrence of a direct or a sequential capture
mechanism would imply a different behavior of the tail of the
photodissociation cross section (see Eqs.(\ref{gamseq3}) and
(\ref{dir2})), and therefore it would lead to a different reaction
rate. This is in fact observed in the schematic case in the upper part
of Fig.~\ref{scheme}, where an example similar to the triple alpha
reaction showed a model dependent rate varying by several orders of
magnitude at low temperatures.

\subsection{The photodissociation cross section}

As mentioned above, for temperatures higher than 0.1 GK the direct and
sequential descriptions provide similar results. Therefore, for
$T>0.1$ GK, the simplest choice for the photodissociation cross
section of $^{12}$C is the sequential expression given by
Eq.(\ref{bwseq}). We use the same values for the energies and widths
of the different states in $^{12}$C and $^8$Be as the ones specified
in \cite{ang99}. The energy dependence of the widths is then as given
by Eq.(\ref{gamwidth}) (with $\lambda=2$) for $\Gamma_\gamma(E)$, and
as given by Eq.(\ref{gamseq3}) (with the $b$-parameter (\ref{babc}))
for $\Gamma_{ab,c}(E)$.

\begin{figure}[t]
\psfig{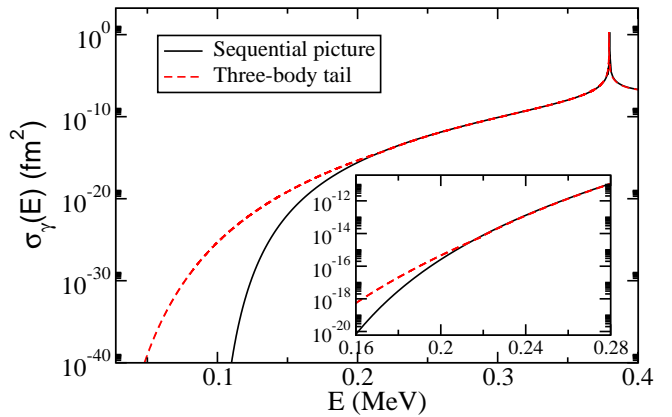}
\caption{(color online) 
Photodissociation cross section as given in Eq.(\ref{bwseq}) for the 
$\mbox{$^{12}$C}+\gamma \rightarrow \alpha+\alpha+\alpha$
reaction as a function of the three-body energy $E$. The parameters are
as given in \cite{ang99}. The solid line has been obtained assuming a sequential
mechanism, using the widths as given in (\ref{gamseq3}) and (\ref{babc}). The dashed line 
corresponds to a direct mechanism, using the widths as given in (\ref{dir2}) and
(\ref{bdir}).  The inner part is a zoom of the cross section in the matching region.}
 \label{fig1}
\end{figure}

If we now assume that the same energy dependence, corresponding to a
sequential mechanism, is still valid for low energies, we then get the
photodissociation cross section shown by the solid line in
Fig.~\ref{fig1}.  The peak at about 0.38 MeV corresponds to the Hoyle
state.

The other possibility is that at low energies the photodissociation
cross section be dominated by a direct process. In this case the tail
of the cross section would be given by Eq.(\ref{dir2}), where the
$b$-parameter is now as in Eq.(\ref{bdir}). Then Eq.(\ref{bweq}) leads
to the cross section shown by the dashed line in Fig.~\ref{fig1}. The
tail of the photodissociation cross section matches well with the one
in the sequential picture (solid line) at about 0.2 MeV, as seen in
the inner part of the figure.

As we can see, for three-body energies smaller than 0.2 MeV the dashed
and solid curves are very different.  Although still small in
absolute values, the cross section in the direct picture (dashed line)
can be orders of magnitude bigger than the one in the sequential
picture (solid line).  This difference is necessarily giving rise to different
reaction rates at small temperatures. How sizable the difference is
between the reaction rates is the topic of the following subsection.

\subsection{Reaction rates}

\begin{figure}[t]
\psfig{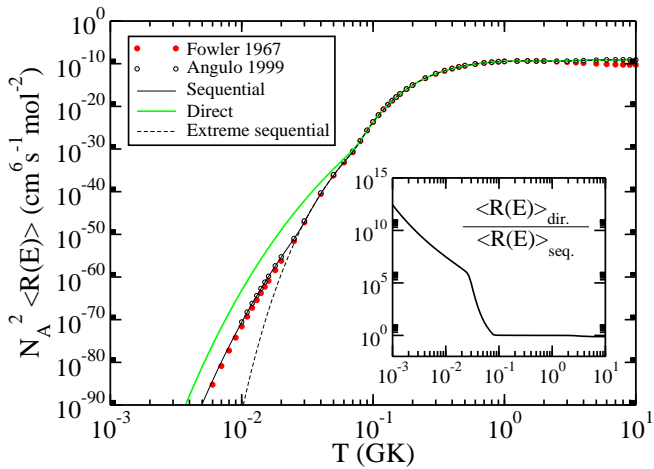}
\caption{(color online) Reaction rate for the triple $\alpha$ process. The solid and open circles are
the results from \cite{fow67} (Fowler 1967) and \cite{ang99} (Angulo 1999), respectively. The thick 
solid curve is the calculation 
as in Eq.(\ref{eq5}) assuming a direct capture process at very low energies. The thin solid line is 
the calculation in the sequential picture (\ref{fullseq}). The dashed line has been obtained in the 
extreme sequential approximation (\ref{seq}).
The inset shows the ratio between the thick solid rate (direct capture assumption at
low energies) and the thin solid line (sequential capture assumption at low energies).  }
 \label{fig2}
\end{figure}

In Fig.~\ref{fig2} the solid and open circles are the reaction rates
for the triple alpha reaction given by Fowler et al. in
Ref.\cite{fow67} and the NACRE compilation \cite{ang99},
respectively. The main difference between these rates is found at high
temperatures, where the result from \cite{fow67} is below the one of
NACRE \cite{ang99}.  This is due to the fact that while the effects of
the first 2$^+$ resonance are included in \cite{ang99}, they are
omitted in \cite{fow67}. At low temperatures, both calculations,
assuming both a fully sequential capture mechanism, provide
essentially the same result.  In the calculations shown in this
section the lowest $2^+$ resonance in $^{12}$C has been included using
the resonance parameters given in \cite{ang99}.

In the figure, the thin dashed line gives the reaction rate obtained
in the extreme sequential model, i.e., when Eq.({\ref{bwseq}}) (solid
curve in Fig.~\ref{fig1}) is inserted into Eq.({\ref{seq}}).  This
approximation amounts to cutting the tail of the cross section for
energies smaller than the one of the intermediate 0$^+$ state in
$^8$Be.  This strong approximation gives rise to a reaction rate in
good agreement with the results in \cite{fow67} and \cite{ang99} at
high temperatures, while for temperatures smaller than $\sim 0.03$ GK
it clearly underestimates the rate. This is consistent with
Fig.~\ref{wths}, where we see that a decreasing width of the three-body
resonance increases the range of temperatures where the sequential and
the extreme sequential calculations agree.

A full sequential description of the process requires the use of
Eq.(\ref{fullseq}).  Compared to the extreme sequential limit, this
calculation includes now the low energy tails of the cross sections
(\ref{eq7}) and (\ref{bwseq}), which are given by (\ref{gamseq4}) and
(\ref{gamseq3}), respectively. This is done in \cite{ang99}, and
since we are using the same resonance parameters, we of course recover
the same result (thin solid line in the figure).

\begin{table}
\begin{tabular}{|c|ccc|} \hline
T (GK)    &  \multicolumn{3}{c|}{Reaction Rate (cm$^6$ s$^{-1}$mol$^{-2}$) } \\ \hline
      &    Direct  & Sequential & Ref.\cite{ang99} (adopted)\\ \hline
0.001 &  $4.50\times 10^{-144}$  &  $1.64\times 10^{-156}$ &  --- \\
0.002 &  $4.97\times 10^{-113}$  &  $1.14\times 10^{-123}$ &  --- \\
0.003 &  $5.47\times 10^{-98} $  &  $1.02\times 10^{-107}$ &  --- \\
0.004 &  $1.64\times 10^{-88} $  &  $1.20\times 10^{-97} $ &  --- \\
0.005 &  $8.99\times 10^{-82} $  &  $1.77\times 10^{-90} $ &  --- \\
0.006 &  $1.23\times 10^{-76} $  &  $5.26\times 10^{-85} $ &  --- \\
0.007 &  $1.54\times 10^{-72} $  &  $1.25\times 10^{-80} $ &  --- \\
0.008 &  $3.71\times 10^{-69} $  &  $5.12\times 10^{-77} $ &  --- \\
0.009 &  $2.66\times 10^{-66} $  &  $5.83\times 10^{-74} $ &  --- \\
0.01  &  $7.67\times 10^{-64} $  &  $2.52\times 10^{-71} $ &  $2.93\times 10^{-71}$\\
0.02  &  $1.09\times 10^{-49} $  &  $4.65\times 10^{-56} $ &  $5.45\times 10^{-56}$\\
0.03  &  $7.90\times 10^{-43} $  &  $8.20\times 10^{-48} $ &  $1.46\times 10^{-47}$\\
0.04  &  $1.71\times 10^{-38} $  &  $3.16\times 10^{-41} $ &  $5.31\times 10^{-41}$\\
0.05  &  $2.15\times 10^{-35} $  &  $6.68\times 10^{-37} $ &  $1.04\times 10^{-36}$\\
0.06  &  $5.19\times 10^{-33} $  &  $8.16\times 10^{-34} $ &  $1.20\times 10^{-33}$\\
0.07  &  $4.92\times 10^{-31} $  &  $2.18\times 10^{-31} $ &  $3.00\times 10^{-31}$\\
0.08  &  $8.84\times 10^{-29} $  &  $7.82\times 10^{-29} $ &  $9.68\times 10^{-29}$\\
0.09  &  $2.22\times 10^{-26} $  &  $2.12\times 10^{-26} $ &  $2.52\times 10^{-26}$\\
0.10  &  $2.11\times 10^{-24} $  &  $2.04\times 10^{-24} $ &  $2.38\times 10^{-24}$\\ \hline
\end{tabular}
\caption{Computed reaction rates for the $\alpha+\alpha+\alpha \rightarrow \mbox{$^{12}$C}+\gamma$ 
reaction assuming a direct capture for the low energy tail of
$\sigma_\gamma$ (second column) and assuming a fully sequential process (third column). The fourth
column gives the adopted reaction rate given in the NACRE compilation \cite{ang99}.}
\label{tab1}
\end{table}

However, when the low energy tail in $\sigma_\gamma$ is assumed to be
given by (\ref{dir2}), which corresponds to a direct capture (dashed
curve in Fig.~\ref{fig1}), and the $\sigma_\gamma$ cross section
(\ref{bweq}) is inserted into Eq.(\ref{eq5}), we then get the reaction
rate shown by the thick solid line in Fig.~\ref{fig2}. As we can see,
for temperatures smaller than $\sim 0.07$ GK, the reaction rate
obtained assuming a direct capture at very low energies is several
orders of magnitude bigger than when the sequential capture is
assumed. This is appreciated more quantitatively in the inset of the
figure, where we show the ratio between both reaction rates, direct
and sequential. We can immediately see that the ratio increases when
decreasing the temperature. For a temperature of $10^{-3}$ GK we have
obtained a reaction rate for the direct capture about 12 orders of
magnitude bigger than in the sequential picture. This difference
reduces to about 7 orders of magnitude for T=0.01 GK.

In table~\ref{tab1} we give the computed reaction rates in the direct
picture (second column) and in the sequential picture (third column)
for temperatures from $10^{-3}$ GK to 0.1 GK. In Ref.\cite{ang99} a
lower limit, a higher limit, and an adopted value are given for this
reaction rate. Our results in the sequential approach lie in between
the two limits given in \cite{ang99}, although closer to the lower
one. In the last column of table~\ref{tab1} we give the reaction rates
quoted in Ref.\cite{ang99} as adopted rate.

It is important to note that the computed rates are the ones obtained
in the limiting cases of a fully direct or a fully sequential
description in the very low energy region. If both processes compete,
the computed reaction rate would then be found in between the thin and
thick solid curves in Fig.~\ref{fig2}, which can be taken as the upper
and lower limits to the true reaction rate. In any case, even if the
process is considered to be fully direct, the increase in the reaction
rate compared to the NACRE result at T=0.01 GK is of only 7 orders of
magnitude (see table~\ref{tab1}), and thus far smaller then the 20
orders of magnitude obtained in \cite{oga09}.

We emphasize that the calculation in the direct picture has been
made with the same value of $\Gamma_\gamma$ ($=3.7\times 10^{-3}$ eV) 
as in the sequential picture.  A change in $\Gamma_\gamma$ implies
precisely the same change in the reaction rate, as seen immediately
from Eqs.(\ref{bweq}) and (\ref{eq5}).

\section{The $^9$Be rate}

Let us consider the $\alpha+\alpha+n \rightarrow \mbox{$^9$Be} +
\gamma$ reaction where the principles are similar to the case of
$^{12}$C.  The properties of the low energy spectrum of $^9$Be will
determine the reaction rate at low temperatures.  As for $^{12}$C, the
presence of the internal two-body subsystem, $^{8}$Be with the very
narrow $0^+$ resonance at 0.092 MeV, suggests a sequential description
of the capture process as the most appropriate model
\cite{ang99,sum02}.

The main difference compared to $^{12}$C is that now $^9$Be does not
show a low-lying and narrow three-body resonance like the Hoyle
state. Instead, $^9$Be has a low lying $1/2^+$ resonance in the
vicinity of 0.11 MeV \cite{ajz88}, only slightly above the 0.092 MeV
of the 0$^+$ resonance in $^8$Be. The width of the 1/2$^+$ state is
estimated to be around 0.1 MeV, although 0.2 MeV is not numerically
excluded \cite{gar10}. As a consequence the photodissociation cross
section shows a relatively broad peak at a three-body energy of around
0.11 MeV, in such a way that $\sigma_\gamma$ is not negligible in the
vicinity of the two-body resonance energy $E_r$=0.092 MeV.  The system
under investigation is now similar to the schematic case shown in the
lower part of Fig.~\ref{schemb}.

\subsection{The photodissociation cross section}

\begin{figure}
\epsfig{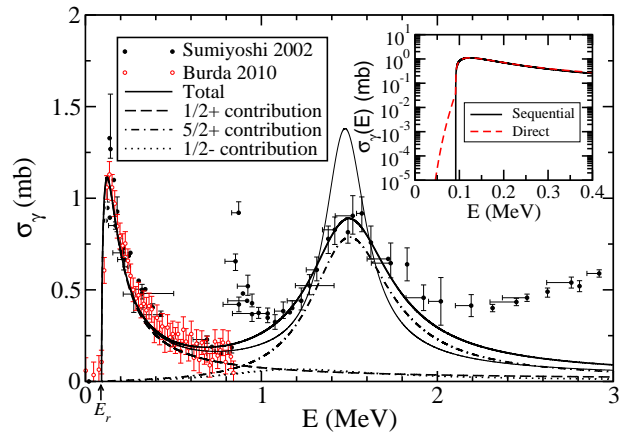}
\caption{(color online) Photodissociation cross section for $^9$Be as a function of the three-body energy $E$. 
The experimental data are from \cite{sum02} (Sumiyoshi 2002, solid circles) and from \cite{bur10} 
(Burda 2010, open circles). 
The dashed, dot-dashed, and dotted curves correspond to the contribution of the 1/2$^+$, 
5/2$^+$, and 1/2$^-$ states in $^9$Be, respectively. The thick solid line gives the sum of the three 
contributions. The 5/2$^+$ resonance is fitted as in \cite{sum02}. The solid thin line is the total 
cross section when the 5/2$^+$ resonance is parametrized as in \cite{ang99}.
The vertical arrow in the $x$-axis indicates the energy of the $0^+$ resonance in $^8$Be. The inner part 
is a zoom of the low energy region of the cross section, where the solid and dashed lines are the cross
section when the sequential and direct pictures are assumed in the low energy region, respectively.}
\label{fig3}
\end{figure}

The experimental photodissociation cross section of $^9$Be can be
found in \cite{sum02}, and it is shown in Fig.~\ref{fig3} by the solid
circles as a function of the three-body energy $E$. Very recently new
data for the peak corresponding to the 1/2$^+$ resonance in $^9$Be
have been published in \cite{bur10}, and they are shown by the open
circles in the figure.  The experimental data show that for three-body
energies below the $0^+$ resonance at 0.092 MeV in $^8$Be, the cross
section essentially vanishes, or at least it is extremely small.  The
energy of the two-body 0$^+$ resonance is indicated by the arrow on
the $x$-axis.

This fact supports the assumption that the intermediate $0^+$ state in
$^8$Be is actually populated in the process, and therefore the
sequential description appears to be appropriate. In fact, the energy
dependence of the cross section (\ref{bwseq}) for sequential decay of
$^9$Be into $^8$Be plus a neutron is governed by (\ref{gamseq1}),
which implies that $\sigma_\gamma$ vanishes when $E^\prime=0$, i.e.,
when $E=E_r$. Furthermore, for high temperatures, a pure three-body
calculation of the reaction rate, without any additional assumption
about the reaction mechanism, agrees reasonably well with the one
obtained in the sequential picture \cite{die10}.

As for $^{12}$C, the behavior of $\sigma_\gamma$ at very low energies,
where again $\sigma_\gamma$ takes very small values, will determine the value of the
reaction rate at very low temperatures, and this rate could change
depending on which model, direct or sequential, is assumed for the
capture mechanism at such low energies.  To investigate this issue we
proceed as for $^{12}$C.  At high temperatures the three-body and
sequential pictures provide similar results (as we can see in the
lower part of Fig.~\ref{schemb}).  For energies higher than $E_r$ we
then start by taking the simple parametrization of the cross section
used in the sequential description.  In particular, we again choose
the resonance parameters given in \cite{ang99}, and the $1/2^+$,
$5/2^+$, $1/2^-$ resonances in $^9$Be are included in the calculation.
This gives rise to the cross section shown in Fig.~\ref{fig3} by the
thin solid line, where the energy dependence given by
Eq.(\ref{gamwidth}) with $\lambda=1$, and Eq.(\ref{gamseq1}), has been
used for all the resonances.

The peak corresponding to the 5/2$^+$ resonance is clearly
overestimated. To solve this problem, we have taken for this
particular $^9$Be resonance the energy and width used in \cite{sum02},
which gives a much better agreement with the experimental cross
section, as shown by the thick solid line. In any case, the properties
of the 5/2$^+$ resonance play a  minor role in the low temperature
behavior of the reaction rate.  The relevant resonance in this
temperature region is the $1/2^+$ state, for which similar parameters
are employed in \cite{ang99} and \cite{sum02}. In Fig.~\ref{fig3} the
dashed, dot-dashed, and dotted curves show the contribution to the
total cross section (thick solid curve) from the $1/2^+$, $5/2^+$, and
$1/2^-$ resonances, respectively.

In the inner part of Fig.~\ref{fig3} we show a zoom of the low energy
region of the photodissociation cross section.  When the sequential
picture is assumed, the cross section takes the form (\ref{bwseq})
with $\Gamma_{ab,c}$ given by (\ref{gamseq1}), and then
$\sigma_\gamma$ vanishes for $E=E_r$, as shown by the solid line in
the inset of the figure. However, if for such low energies we assume a
direct mechanism, due to the Coulomb repulsion of the two alpha
particles, the low energy behavior of the cross section takes the form
(\ref{dir2}), and the cross section behaves as shown by the dashed
line in the inset of Fig.~\ref{fig3}.

\subsection{Reaction rates}

\begin{figure}
\epsfig{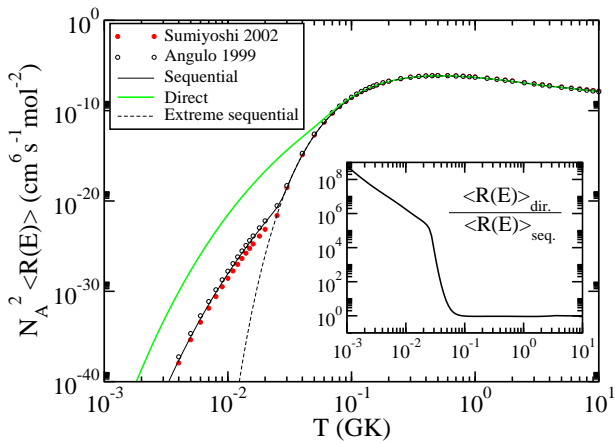}
\caption{(color online) Reaction rate for the reaction 
$\alpha+\alpha+n \rightarrow \mbox{$^9$Be}+\gamma$. The solid and open circles are
the results from \cite{sum02} (Sumiyoshi 2002)  and \cite{ang99} (Angulo 1999), respectively. 
The dashed line has been obtained in the
extreme sequential approximation (\ref{seq}). The thin solid line is the calculation in the sequential
picture (\ref{fullseq}). The thick solid curve is the calculation (\ref{eq5}) assuming a direct
capture process at very low energies (dashed curve in the inset of Fig.~\ref{fig3}). 
The inset shows the ratio between the thick solid rate (direct capture assumption at
low energies) and the thin solid line (sequential capture assumption at low energies).  }
\label{fig4}
\end{figure}

The solid and open circles in Fig.~\ref{fig4} are the reaction rates
for the $\alpha+\alpha+n \rightarrow \mbox{$^9$Be}+\gamma$ process
given by Sumiyoshi et al. in Ref.\cite{sum02} and the NACRE
compilation \cite{ang99}, respectively. They basically agree in the
whole temperature range shown in the figure, although for low
temperatures the results from \cite{sum02} are slightly below the ones
in \cite{ang99}. This is due to the fact that the treatment of the low
energy tails in the cross section is not exactly the same in both
cases. More precisely, for the 5/2$^+$ resonance the widths involved
in Eqs.(\ref{eq7}) and (\ref{bwseq}) are taken as constant in
\cite{ang99}, while in \cite{sum02} they are assumed to be energy
dependent. In the present work we use the same energy dependence as
for the 1/2$^+$ resonance, and therefore our computed reaction rates
in the sequential picture at very low temperatures should also be a bit
smaller than the ones given in \cite{ang99}.

If we now take the cross section given by the thick solid line in
Fig.~\ref{fig3} and use it in Eq.(\ref{seq}), we then get the reaction
rate in the extreme sequential picture, which is shown by the dashed
line in Fig.~\ref{fig4}. As expected, this crude approximation fails
completely at low temperatures. As we already know, a correct
sequential calculation requires use of Eq.(\ref{fullseq}), which
implies use of the cross section (\ref{eq7}) and consequently
inclusion of the energy tail (\ref{gamseq4}) coming from the decay of
the $0^+$ resonance in $^8$Be. When this is done we obtain the rate
given by the thin solid line in Fig.~\ref{fig4}, which, as expected,
are slightly below the results given in \cite{ang99} (open circles) at very low
temperatures.

\begin{table}
\begin{tabular}{|c|ccc|} \hline
T (GK)    &  \multicolumn{3}{c|}{Reaction Rate (cm$^6$ s$^{-1}$mol$^{-2}$) } \\ \hline
      &    Direct  & Sequential & Ref.\cite{ang99} (adopted)\\ \hline
0.001 &  $2.35\times 10^{-50}$  &  $9.00\times 10^{-60}$ &  $3.90\times 10^{-59}$ \\
0.002 &  $2.25\times 10^{-39}$  &  $7.62\times 10^{-48}$ &  $2.50\times 10^{-47}$ \\
0.003 &  $4.30\times 10^{-34} $  &  $4.76\times 10^{-42}$ &  $1.35\times 10^{-41}$ \\
0.004 &  $8.80\times 10^{-31} $  &  $2.18\times 10^{-38} $ &  $5.58\times 10^{-38}$ \\
0.005 &  $1.94\times 10^{-28} $  &  $8.93\times 10^{-36} $ &  $2.11\times 10^{-35}$ \\
0.006 &  $1.17\times 10^{-26} $  &  $8.84\times 10^{-34} $ &  $1.96\times 10^{-33}$ \\
0.007 &  $3.06\times 10^{-25} $  &  $3.50\times 10^{-32} $ &  $7.39\times 10^{-32}$ \\
0.008 &  $4.48\times 10^{-24} $  &  $7.33\times 10^{-31} $ &  $1.48\times 10^{-30}$ \\
0.009 &  $4.30\times 10^{-23} $  &  $9.66\times 10^{-30} $ &  $1.88\times 10^{-29}$ \\ 
0.01  &  $3.00\times 10^{-22} $  &  $8.96\times 10^{-29} $ &  $1.69\times 10^{-28}$\\
0.02  &  $1.99\times 10^{-17} $  &  $4.10\times 10^{-23} $ &  $6.29\times 10^{-23}$\\
0.03  &  $4.14\times 10^{-15} $  &  $2.85\times 10^{-19} $ &  $5.05\times 10^{-19}$\\
0.04  &  $1.21\times 10^{-13} $  &  $1.20\times 10^{-15} $ &  $1.90\times 10^{-15}$\\
0.05  &  $1.51\times 10^{-12} $  &  $1.76\times 10^{-13} $ &  $2.58\times 10^{-13}$\\
0.06  &  $1.29\times 10^{-11} $  &  $4.66\times 10^{-12} $ &  $6.44\times 10^{-12}$\\
0.07  &  $8.03\times 10^{-11} $  &  $4.64\times 10^{-11} $ &  $6.14\times 10^{-11}$\\
0.08  &  $3.54\times 10^{-10} $  &  $2.52\times 10^{-10} $ &  $3.24\times 10^{-10}$\\
0.09  &  $1.16\times 10^{-9} $  &  $9.17\times 10^{-10} $ &  $1.15\times 10^{-9}$\\
0.10  &  $3.02\times 10^{-9} $  &  $2.53\times 10^{-9} $ &  $3.12\times 10^{-9}$\\ \hline
\end{tabular}
\caption{Computed reaction rates for the $\alpha+\alpha+n \rightarrow \mbox{$^{9}$Be}+\gamma$
reaction assuming a direct capture for the low energy tail of
$\sigma_\gamma$ (second column) and assuming a fully sequential process (third column). The fourth
column gives the adopted reaction rate given in the NACRE compilation \cite{ang99}.}
\label{tab2}
\end{table}

Finally, let us assume that the low energy tail in $\sigma_\gamma$
corresponds to a direct decay mechanism (dashed curve in the inset of
Fig.~\ref{fig3}). In this case, the cross section is given by
(\ref{bweq}), and the low energy tail takes again the form
(\ref{dir2}). The reaction rate is then given by Eq.(\ref{eq5}), and
we obtain the thick solid curve in Fig.~\ref{fig4}.  The reaction rate
computed assuming a direct capture at very low energies and the one
obtained assuming a sequential capture begin to differ for
temperatures smaller than about 0.07 GK.  Very soon the rate obtained
in the direct picture is several orders of magnitude bigger than the
sequential one.  This is better appreciated in the inset of the
figure, where we show the ratio between both reaction rates. For a
temperature of 0.01 GK the reaction rate in the direct picture is
almost 7 orders of magnitude bigger than in the sequential
picture. The general behavior of the reaction rates is similar to the
one obtained for the triple alpha reaction, although due to the
smaller Coulomb repulsion, the absolute values of the rates are now
much bigger.

As for $^{12}$C, we now give in table~\ref{tab2} the computed
reactions rates in the direct picture (second column) and in the
sequential picture (third column) for temperatures from $10^{-3}$ GK
to 0.1 GK. In the fourth column we give the adopted value quoted in
\cite{ang99}. As already mentioned, for very low temperatures, our
computed rates in the sequential picture are below (up to more than a
factor of 4 for $T=0.001$ GK) the rates given in \cite{ang99}. In
fact, our values are even below the lower limit for the rate given in
\cite{ang99}. When a constant value is taken for the different
$\Gamma$'s involved in the decay of the 5/2$^+$ resonance, a rate
within the limits given in \cite{ang99} is obtained.

The comments made when discussing the reaction rates for the triple
alpha reaction are still valid here. First, the computed rates are the
limiting cases of a fully direct process and a fully sequential
process in the very low energy region. If both processes compete, the
computed reaction rate would be found in between the thin and thick
solid curves in Fig.~\ref{fig4}. And second, the calculation in the
direct picture has been made assuming that $\Gamma_\gamma(E)$, given
by Eq.(\ref{gamwidth}), is the same as in the sequential picture, with
$\Gamma_\gamma=0.51$ eV. A change in $\Gamma_\gamma$ would imply the
same change in the reaction rate.

\section{Summary and conclusions}

In this work we investigate how different descriptions of the
photodissociation cross sections at very low energies change the
reaction rates for radiative three-body capture processes.  We focus
on the reaction rates at the very low temperatures relevant for the
nucleosynthesis of elements in the core of a star.  More precisely, we
consider the radiative capture of three particles (nuclei, neutrons,
protons\ldots) into a bound nucleus plus a photon.  We investigate how
the reaction rate for such a process changes at low temperatures
depending on the capture mechanism, i.e. either sequentially through
an intermediate two-body state or directly.

We first establish notation and definitions, and in particular we
specify the formulae for sequential and direct reaction rates for
three arbitrary particles. These expressions rely heavily on the
energy dependence of the photodissociation cross section.  We have
previously computed rates and cross sections in a full three-body
computation without assumptions of reaction mechanism.  In these
computations the shape of the cross sections around resonance
positions are computed as well, except at very low energies where the
discretization is insufficient to characterize the resonance shapes,
and a specific form has to be chosen.  These low energies are crucial in
the present work where we assume a simple Breit-Wigner form for the
photodissociation cross section but with energy dependent width as in
$R$-matrix analyses. This is not a severe limitation since the
overall conclusions are general and very robust.

We compute rates as functions of the involved parameters for both two-
and three-body resonances. We study the dependence on reaction
mechanisms for different resonance positions and widths for different
combinations of charges.  The different rates almost coincide for
temperatures somewhat larger than the three-body resonance energy.
The direct decay only depends on position and width of the three-body
resonance entering through the photodissociation cross section, and
thus not on any two-body substructure. On the other hand, the
sequential decay mechanism strongly depends on the relative position
of the intermediate two-body resonance. If the two-body resonance is
above the three-body resonance and the widths are substantially
smaller than the energy difference between these resonances the
sequential rate reproduces almost the extreme limit of zero widths
where the rate expressions are very simple.

When the two and three-body energies exchange positions much larger
rates are found and the sequential and direct rates apparently becomes
more similar. However, they still deviate by many orders of magnitude
strongly depending on charge and, without Coulomb potentials, angular
momentum.  In all cases the sequential rate is substantially smaller
than the direct rate, except when the two and three-body energies are
fairly close to each other and all particles are charged.

We turn to practical and realistic estimates of astrophysical reaction
rates.  We compute rates for the $\alpha+\alpha+\alpha\rightarrow
\mbox{$^{12}$C}+\gamma$ and the $\alpha+\alpha+n\rightarrow
\mbox{$^9$Be}+\gamma$ reactions.  Three-body calculations \cite{die10}
show that at high temperatures the usual sequential description
\cite{fow67,ang99,sum02} is appropriate.  We focus on the reaction
rates at low temperatures, where due to the extremely small energies,
clearly below the energy of the available intermediate two-body state,
the sequential picture is questionable.  Therefore, for high
temperatures we take one of the sequential parametrizations of the
photodissociation cross section available in the literature. For the
low energy tail we consider the possibility of a direct process for
which the penetration factors are known for both Coulomb and
centrifugal barrier potentials.

We find that a direct description of the low energy tail of the
photodissociation cross section enhances the reaction rates at low
temperatures by several orders of magnitude compared to the sequential
description. For a temperature of 0.01 GK the difference is of about
7 orders of magnitude for the two reactions investigated.
These reaction rates in the direct and sequential
pictures are the upper and lower limits of the rates. Therefore the
result obtained for the triple alpha reaction in \cite{oga09} is not
consistent with our calculation.

In conclusion, comparison between sequential and direct capture
mechanisms reveal large differences at temperatures below the
three-body resonance energy. These temperatures are important for
the capture processes producing light nuclei in aging stars.  The
traditional use of the sequential models is insufficient and the
direct reaction mechanism should be employed for these temperatures.
Usually this leads to a substantial enhancement of the rates, but the
precise increase depends on the characteristics of the most important
resonances.  Thus, individual calculations are necessary for each
process.

\begin{acknowledgement}
This work was partly supported by funds provided by DGI of MEC (Spain)
under contract No.  FIS2008-01301. One of us (R.D.) acknowledges
support by a Ph.D. I3P grant from CSIC and the European Social Fund.
\end{acknowledgement}

\end{document}